\documentclass[12pt]{article} 
\usepackage{graphicx}
\usepackage{epsfig}
\usepackage{amsbsy}

\begin{document}
\baselineskip 10mm

\centerline{\large \bf Stone–Wales Transformation Paths in Fullerene C$_{60}$}

\vskip 6mm

\centerline{A. I. Podlivaev and L. A. Openov$^{*}$}

\vskip 4mm

\centerline{\it Moscow Engineering Physics Institute (State
University), 115409 Moscow, Russia}

\vskip 2mm

$^{*}$ E-mail: opn@supercon.mephi.ru

\vskip 8mm

\centerline{\bf ABSTRACT}

The mechanisms of formation of a metastable defect isomer of
fullerene C$_{60}$ due to the Stone-Wales transformation are
theoretically studied. It is demonstrated that the paths of the
{}''dynamic'' Stone-Wales transformation at a high (sufficient for
overcoming potential barriers) temperature can differ from the two
{}''adiabatic'' transformation paths discussed in the literature. This
behavior is due to the presence of a great near-flat segment of
the potential-energy surface in the neighborhood of metastable
states. Besides, the sequence of rupture and formation of
interatomic bonds is other than that in the case of the adiabatic
transformation.

\newpage

Various models of fullerene growth have been proposed [1]. Because
a sphere-shaped cluster C$_{60}$ is formed at a high temperature
under actual operating conditions, its structure can be
significantly different from the structure of ideal
buckminsterfullerene with $I_{h}$ icosahedral symmetry. On the
{}''surface'' of buckminsterfullerene, carbon atoms are arranged
at the vertexes of 20 hexagons and 12 pentagons, which are
isolated from each other. Therefore, regardless of the mechanism
of formation of such a spheroidal cluster from graphite fragments
and/or carbon dimers, the question arises of the paths of its
evolution into an equilibrium configuration of
buckminsterfullerene, that is, the question of the mechanisms of
defect annealing that reduces the potential energy $E_{pot}$ of
the cluster.

Defect annealing occurs through local rearrangements of C-C bonds
and is the reverse process for defect formation. The so called
Stone-Wales transformation [2], which consists in the
rearrangement of two C-C bonds in buckminsterfullerene (see Fig.
1), results in a defect isomer closest to buckminsterfullerene in
terms of energy. This metastable isomer exhibits $C_{2v}$ symmetry
and contains two pairs of pentagons with common sides. Among a
great number of other isomers, this isomer stands out as the last
segment (before buckminsterfullerene) in the chain of sequentially
decreasing energies of a C$_{60}$ cluster on defect annealing [3].

In this context, attention was focused on studying changes in the
mutual arrangement of atoms in the Stone-Wales transformation and
on determining the height $U$ of a minimum energy barrier to this
transformation [4-10]. Two different transformation paths were
considered: (I) the rotation of a C-C bond shared by two hexagons
through 90$^{0}$ so that all of the atoms remained on the cluster
{}''surface'' and (II) the rupture of a C-C bond shared by
pentagon and hexagon; thereafter, one of the atoms initially rose
above the cluster {}''surface'' and then relaxed to form new
bonds. In accordance with the calculations of $E_{pot}$ as a
function of atomic coordinates $\{{\bf R}_{i}\}$ performed using
the density functional theory [5 - 10], the barrier height for
path I was $U_{I}$ = 6 - 8 eV, whereas $U_{II}$ = 6.64 - 7.6 eV
for path II. A considerable scatter in the values of $U_{I}$ and
$U_{II}$ does not allow one to determine unambiguously which of
the two paths exhibits a lower barrier and, consequently, by which
mechanism the Stone-Wales transformation occurs. Even with the use
of the same exchange-correlation functional and equal
basis-function sets (or equal cutoff energies in a plane-wave
basis) for calculating $U_{I}$ and $U_{II}$, different results
were obtained: $U_{I}=U_{II}$ = 7.27 eV [10], $U_{I}$ = 6.30 eV
$<$ $U_{II}$ = 6.64 eV [9], and $U_{I}$ = 8.1 eV $>$ $U_{II}$ =
7.6 eV [6, 7].

Note that the determination of the height of a barrier and the
type of a transition state based on the analysis of the shape of a
potential-energy surface implies an adiabatic transition of the
cluster from one state to another. That is, in such a transition,
one or more properly chosen interatomic distances change in
accordance with a certain law with a small step along the reaction
coordinate, whereas relaxation to a minimum (with consideration
for imposed restrictions) total energy is performed at each step
in all of the other degrees of freedom. Physically, an adiabatic
transition corresponds to a forced deformation of the cluster at
$T=0$, and it must not occur identically to transitions at a
finite (especially, high) temperature. In this work, we studied
Stone-Wales transformation paths both in an adiabatic transition
and at a high temperature (sufficient for overcoming a potential
barrier) using a molecular dynamic simulation of the {}''life'' of
buckminsterfullerene. A comparison between the results allowed us,
on the one hand, to find common features in the low-temperature
and high temperature buckminsterfullerene metastable isomer
transitions and, on the other hand, to reveal considerable
differences between them.

We used a tight binding potential for calculations. This potential
incorporates the contribution of all the valence electrons (four
electrons from each atom) to the total energy in an explicit form,
and it is actually an $n-$body potential, where $n$ is the number
of atoms in the cluster. Because of this, this potential surpasses
simple but less reliable empirical potentials. Although this
method is not as strict as {\it ab initio} approaches, it is
highly competitive with them in the accuracy of the description of
carbon systems [11]. Moreover, this method considerably
facilitates the simulation of dynamic processes because it is not
resource-intensive. Previously, we used this method for an
analysis of the thermal stability of metastable C$_{8}$ and
C$_{20}$ clusters [12 - 14]. For buckminsterfullerene, it gave the
binding energy $E_{b} = 60E$(C$_{1}$) - $E$(C$_{60}$) = 6.86
eV/atom and the bond lengths $l$ = 1.396 and 1.458 $\AA$, which
are consistent with the experimental values of $E_{b}$ = 6.97 -
7.01 eV/atom [15] and $l$ = 1.402 and 1.462 $\AA$ [16]. The
HOMO-LUMO gap $\Delta$ = 1.62 eV is also consistent with the
experimental value of $\Delta$ = 1.6 - 1.8 eV [17].

Initially, we studied the mechanisms of the buckminsterfullerene
$\rightarrow$ metastable isomer isomerization by adiabatic
transitions. For this purpose, we calculated the "potential
relief" $E_{pot}(\{{\bf R}_{i}\})$ for the C$_{60}$ cluster in the
vicinity of an equilibrium atomic configuration
(buckminsterfullerene) and the closest in energy metastable atomic
configuration (metastable isomer) for the above two Stone-Wales
transformation paths. Figure 2 shows the results of this
calculation. The energy of the metastable isomer is higher than
the energy of buckminsterfullerene by $\Delta E$ = 1.42 eV, which
is consistent with the value of $\Delta E$ = 1.4 - 1.7 eV obtained
by the density functional method [5 - 10]. In both path I and path
II, the dependence of $E_{pot}$ on reaction coordinate $X$
exhibits the only stationary point (a maximum), which is a saddle
point for the $E_{pot}(\{{\bf R}_{i}\})$ surface and determines
the barrier height. According to our calculations, the value of
$U_{I}$ = 6.46 eV is somewhat lower than $U_{II}$ = 6.58 eV.

Figure 3a shows the atomic configuration of the transition state
for path I. It is "symmetric" in accordance with available
published data [6 - 10]. On the contrary, the transition state for
path II is asymmetric (Fig. 3b). One of the atoms is arranged over
the cluster {}''surface'', and it forms two bonds ($sp$
hybridization). In this case, another ({}''surface'') atom has
four neighbors ($sp^{3}$ hybridization). This transition state has
been reported previously [4, 6, 7, 9, 10]. Note that path II
exhibits an almost horizontal segment near a maximum in the
$E_{pot}(X)$ curve. The atomic configuration that corresponds to
the value of $X$ at which the second derivative
$d^{2}E_{pot}(X)/dX^{2}$ exhibits a minimum (Fig. 2, point {\it
4}) is visually similar to the transition state (Fig. 3b). Murry
{\it et al.} [6, 7] stated that this configuration corresponds to
a local minimum of $E_{pot}$ (that is, actually, to another
metastable state). In recent publications [9, 10], no additional
minimum of $E_{pot}$ was reported.

Nevertheless, we found that both of the above Stone-Wales
transformation paths are characterized by barriers with
approximately equal heights: $U_{I}$ = 6.46 eV and $U_{II}$ = 6.58
eV. These data are consistent with first-principle calculations [5
- 10], which gave similar values of $U_{I}$ and $U_{II}$ in a
range of $7 \pm 1$ eV. One might expect that the Stone-Wales
transformation under conditions of a real experiment (that is, due
to thermally activated processes) occurs via both of these paths
with approximately equal probabilities. To test this hypothesis,
we numerically simulated the dynamics of buckminsterfullerene at
$T=$ 4000 - 5000 K and a fixed total energy of the cluster, $E =
E_{kin} + E_{pot}$ = const. The cluster temperature was found from
the equation [18] $k_{B}T(3n-6) = E_{kin}$, where $n=60$, and
$E_{kin}$ is the kinetic energy in the system of the center of
gravity averaged over 10$^{3}$ steps of molecular dynamics (the
single step time $t_{0} = 2.72\cdot10^{–16}$ s was about 0.01 of
the period of the highest frequency mode of cluster vibrations).
On this formulation of the problem, the temperature $T$ is a
measure of energy of relative atomic motion [18]. In this case,
the value of $E_{kin}$ = 30 - 40 eV is sufficient for overcoming
the potential barriers $U_{I,II} \approx$ 6.5 eV (although
$E_{kin} >> U_{I,II}$, buckminsterfullerene isomerization
processes at $T\approx$ 4500 K occur rarely, approximately once a
time of $10^{6} t_{0}$).

We analyzed in detail 45 buckminsterfullerene $\rightarrow$
metastable isomer transitions and vice versa. We found that the
Stone-Wales transformation occurred via path I (or by a visually
very similar mechanism) in the majority of cases (37), see Fig. 4;
it occurred via path II only in three instances, see Fig. 5. Thus,
although both of the paths are almost equally favorable in terms
of energy (see Fig. 2), path I is much more preferable from a
dynamic standpoint. We believe that this is due to a dramatic
difference between the frequency factors $A_{I}$ and $A_{II}$ in
the Arrhenius temperature dependence of the rates of transition
$k_{I,II}=A_{I,II}\exp(–U_{I,II}/k_{B}T)$. Consequently, in the
consideration of Stone-Wales transformation mechanisms, attention
should be focused on an analysis of the frequency factors of
transitions via these two paths rather than on a further increase
in the accuracy of the calculation of barrier heights $U_{I,II}$
(and on determining which of these barriers is lower even by
fractions of eV).

In addition to the transformation of buckminsterfullerene via
paths I and II, we also observed transitions accompanied by the
formation of either a large {}''window'' or three adjacent windows
and the appearance of a "branch" of three atoms, which was
perpendicular to the cluster {}''surface''. All of these
intermediate configurations correspond to potential-energy surface
segments above the saddle points of transitions via paths I and
II, that is, to other transformation paths with higher barriers.
At $T=$ 5000-5500 K, the relative number of these transitions
increases dramatically. They can occasionally result in the
formation of isomers with higher energies (for example, fullerenes
with three or more pairs of adjacent pentagons or nonclassical
fullerenes with tetragons and heptagons).

Note that, in the dynamic transformation of buckminsterfullerene
via path I, as a rule, one or two atoms in the C-C bond that
rotated through 90$^{0}$ (Figs. 1 and 3a) rose above the cluster
{}''surface'' through different heights every time as in the
transformation via path II (but, unlike path II, without the
formation of an $sp^{3}$ configuration). This suggests that there
are a great number of configurations with energies close to the
saddle-point energies of paths I and II. That is, in other words,
a segment in the potential-energy surface between two saddle
points is almost planar. To test this hypothesis, we calculated
$E_{pot}$ as a function of two reaction coordinates $H$ and
$\Theta$ and found that, indeed, the $E_{pot}(H,\Theta)$ function
exhibits a very great radius of curvature along the coordinate $H$
(Fig. 6). Note that a second-order stationary point (characterized
by two imaginary frequencies) occurred between the saddle points
of paths I and II. The energy of this stationary point is 0.56 eV
higher than the energy of the saddle point of path II. This
stationary point corresponds to a configuration in which one of
the atoms of the rotated C-C bond rose over the cluster
{}''surface'' (however, to a lesser degree than in the
transformation via path II).

In conclusion, let us consider the rupture and formation of
interatomic bonds in the dynamic Stone-Wales transformation.
Figure 4 shows a sequence of snapshots of the cluster in the
course of the transformation via path I. It can be seen that the
transformation began with the rupture of a single bond rather than
the concerted rupture of two bonds shared by pentagons and
hexagons, as in an adiabatic case [10]. Physically, this is easy
to understand: a distortion of the ideal buckminsterfullerene
structure due to the thermal vibrations of constituent atoms
caused a symmetry breakdown in the arrangement of these two bonds.
Because of this, one of these bonds became weaker than the other,
and it was ruptured initially. Subsequently, the second bond was
ruptured. Thereafter, a configuration similar to the transition
state in an adiabatic case was formed (Fig. 3a). At the next step,
new bonds were formed (consecutively rather than simultaneously)
to complete the transformation. Note that the overall process
occurred very rapidly and took only $\Delta t_{I} \approx 86$ fs.
The transformation via path II began with the rupture of either a
bond shared by pentagon and hexagon or a bond shared by two
hexagons. For the latter case, Fig. 5 shows a sequence of
snapshots of the C$_{60}$ cluster. The configuration corresponding
to the transition state (Fig. 3b) can be clearly seen in Fig. 5.
The transformation time $\Delta t_{II}\approx 354$ fs is much
longer than the value of $\Delta t_{I}$; this is consistent with a
greater {}''length'' of path II (see Fig. 6).

This work was supported by the U.S. Civilian Research and
Development Foundation (award {}''Research and Education Center
for Basic Investigation of Matter Under Extreme Conditions'').

\vskip 6mm

\centerline{\bf ACKNOWLEDGMENTS}

\vskip 2mm

This study was supported by the CRDF, project {}''Scientific and
Educational Center for Basic Research of Matter in Extremal
States''

\newpage

\includegraphics[width=\hsize]{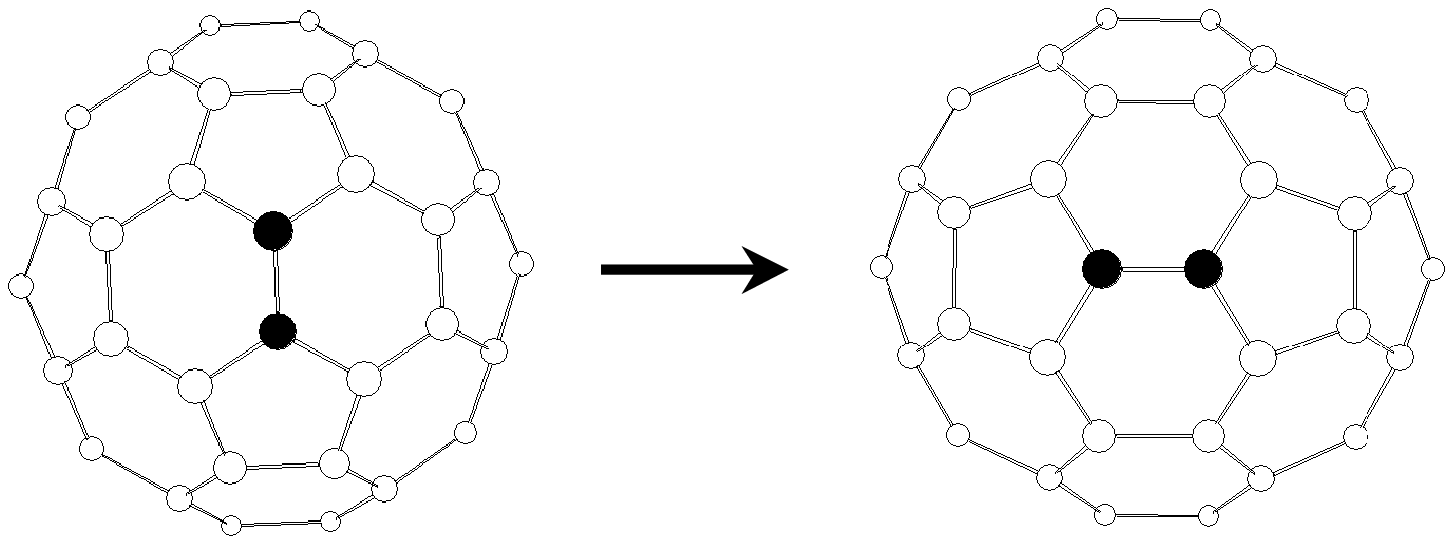}

\vskip 6mm

Fig. 1. Formation of two pairs of pentagons with common sides on
the rearrangement of two C-C bonds in fullerene C$_{60}$
(Stone-Wales transformation). For clarity, atoms in the background
are not shown.

\newpage

\includegraphics[width=0.8\textwidth]{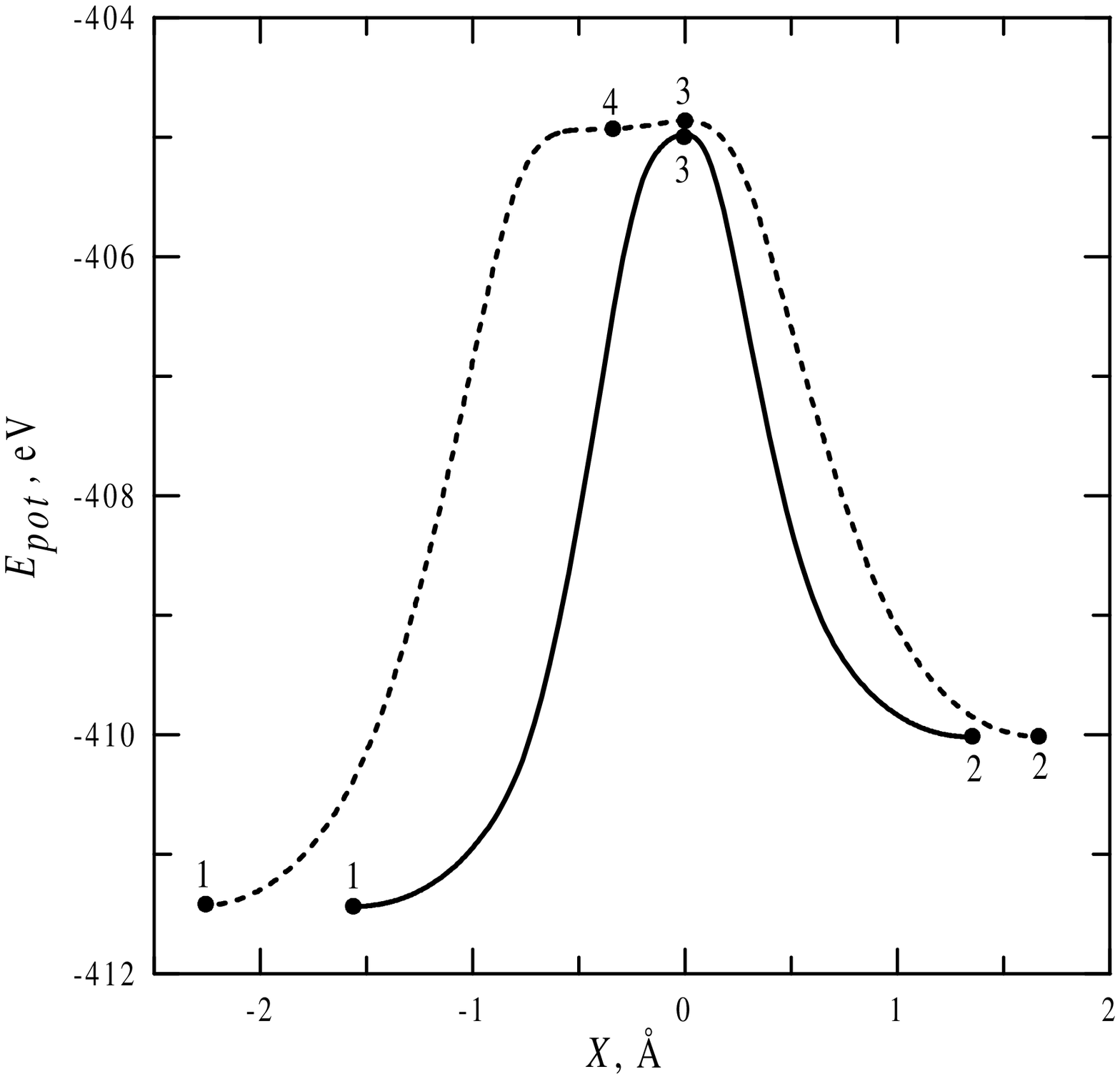}

\vskip 6mm

Fig. 2. Potential energy $E_{pot}$ of the C$_{60}$ cluster vs. the
reaction coordinate $X$ in the neighborhoods of ({\it 1})
equilibrium and ({\it 2}) metastable atomic configurations in the
Stone-Wales transformation (Fig. 1). The energy of 60 isolated
carbon atoms was taken as zero energy. The solid and dashed lines
indicate paths I and II, respectively (see the text); points {\it
3} are $E_{pot}(X)$ maximum points (saddle points for
$E_{pot}(\{{\bf R}_{i}\})$, and {\it 4} is a $d^{2}E_{pot}/dX^{2}$
minimum point for path II. The path length along a trajectory that
passes through the corresponding saddle point in
$(3n–6)$-dimensional space and joins buckminsterfullerene with the
metastable isomer was chosen as the reaction coordinate. $X=0$ at
saddle points.

\newpage

\includegraphics[width=\hsize]{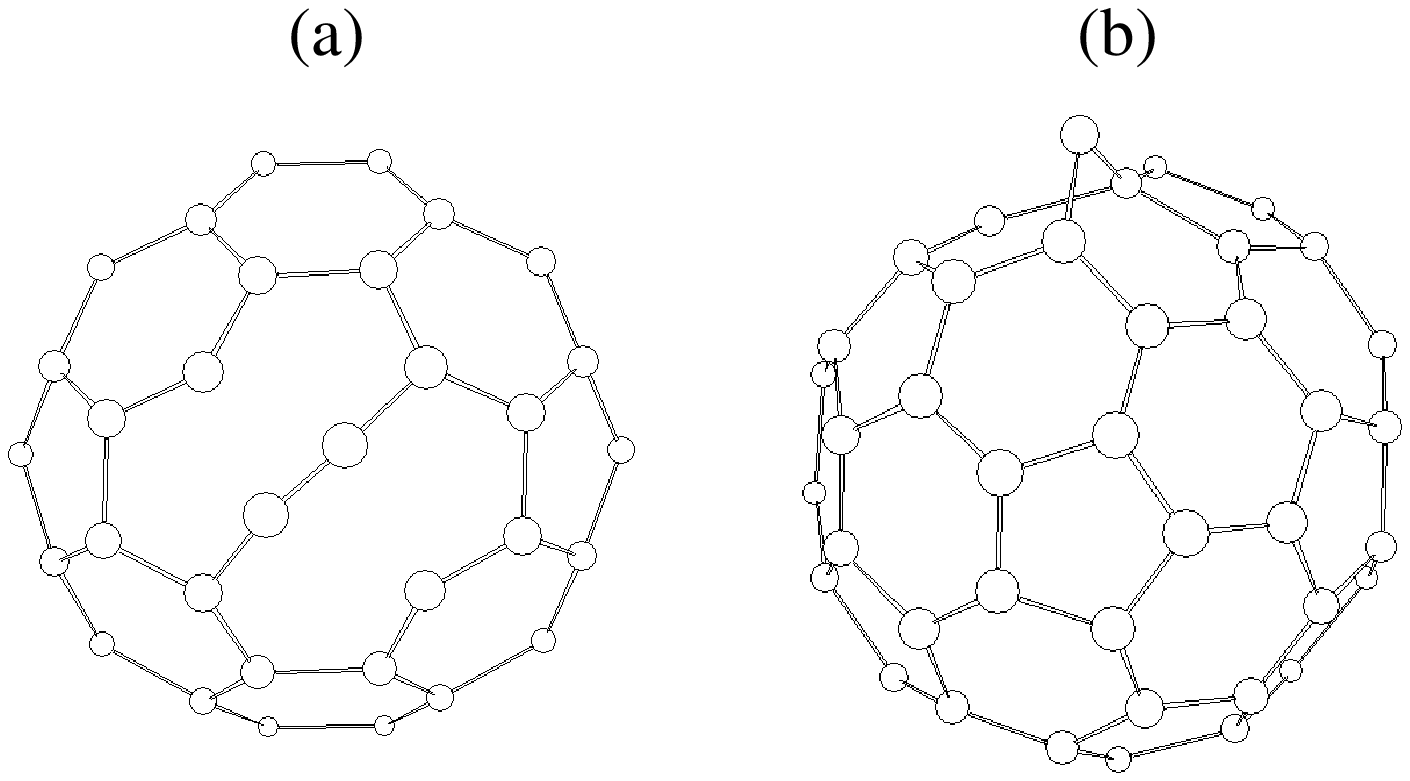}

\vskip 6mm

Fig. 3. Atomic configurations of transition states {\it 3} in Fig.
2 in the Stone-Wales transformation via (a) path I (symmetric
transition state) and (b) path II (asymmetric transition state).
Atoms in the background and C-C bonds more than 2 $\AA$ in length
are not shown.

\newpage

\includegraphics[width=0.6\textwidth]{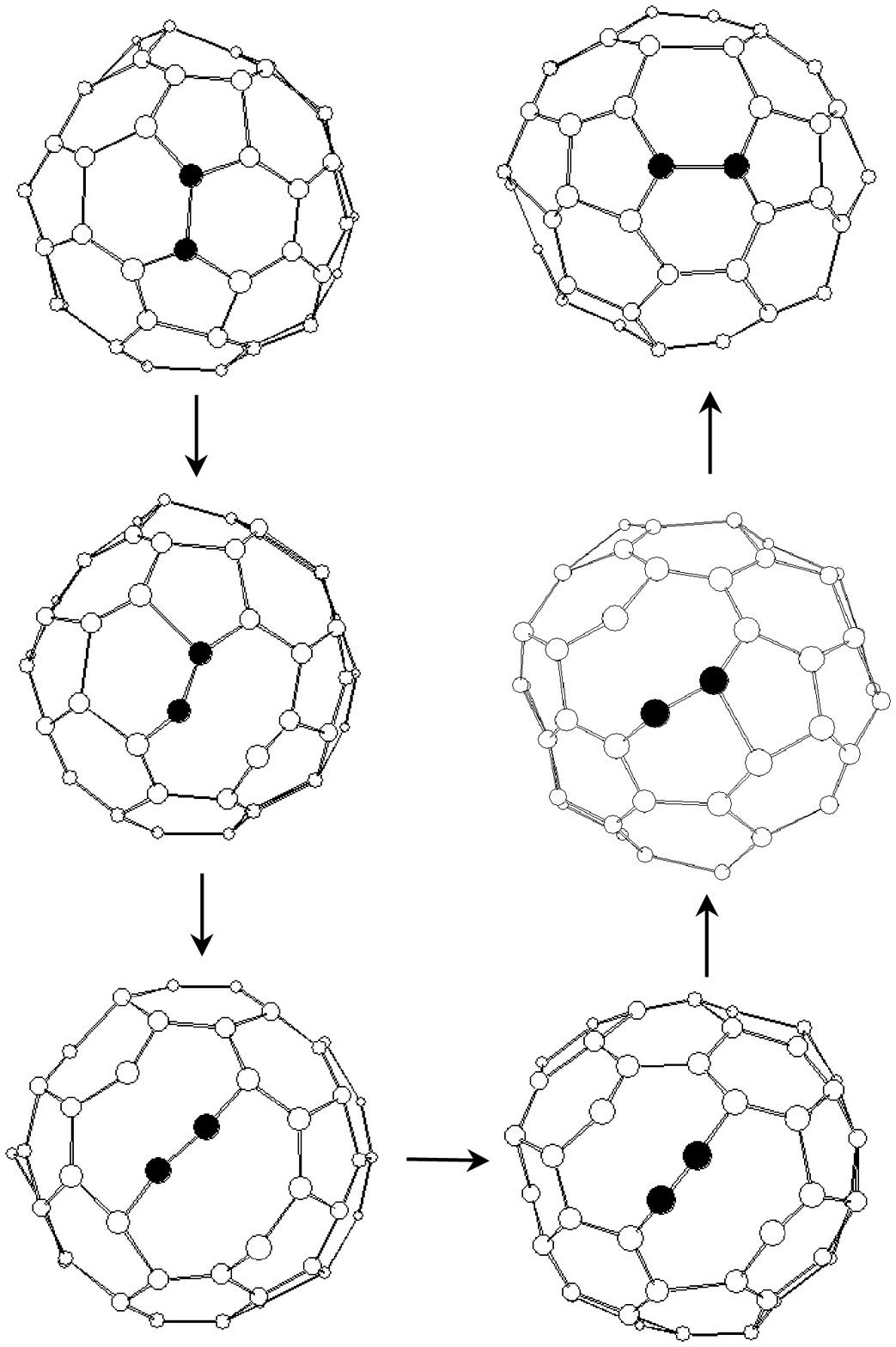}

\vskip 6mm

Fig. 4. Snapshots of the C$_{60}$ cluster in the course of the
dynamic Stone-Wales transformation via path I. Initial temperature
$T_{ini}=4465 \pm 5$ K. The first configuration corresponds to the
time $t^{\prime} = 0.245$ ns, and the subsequent configurations
correspond to the times $t^{\prime} + \Delta t$, where $\Delta t$
= 5.4, 17.7, 69.9, 73.4, and 85.7 fs, respectively. Atoms in the
background and C-C bonds more than 2 $\AA$ in length are not
shown.

\newpage

\includegraphics[width=0.6\textwidth]{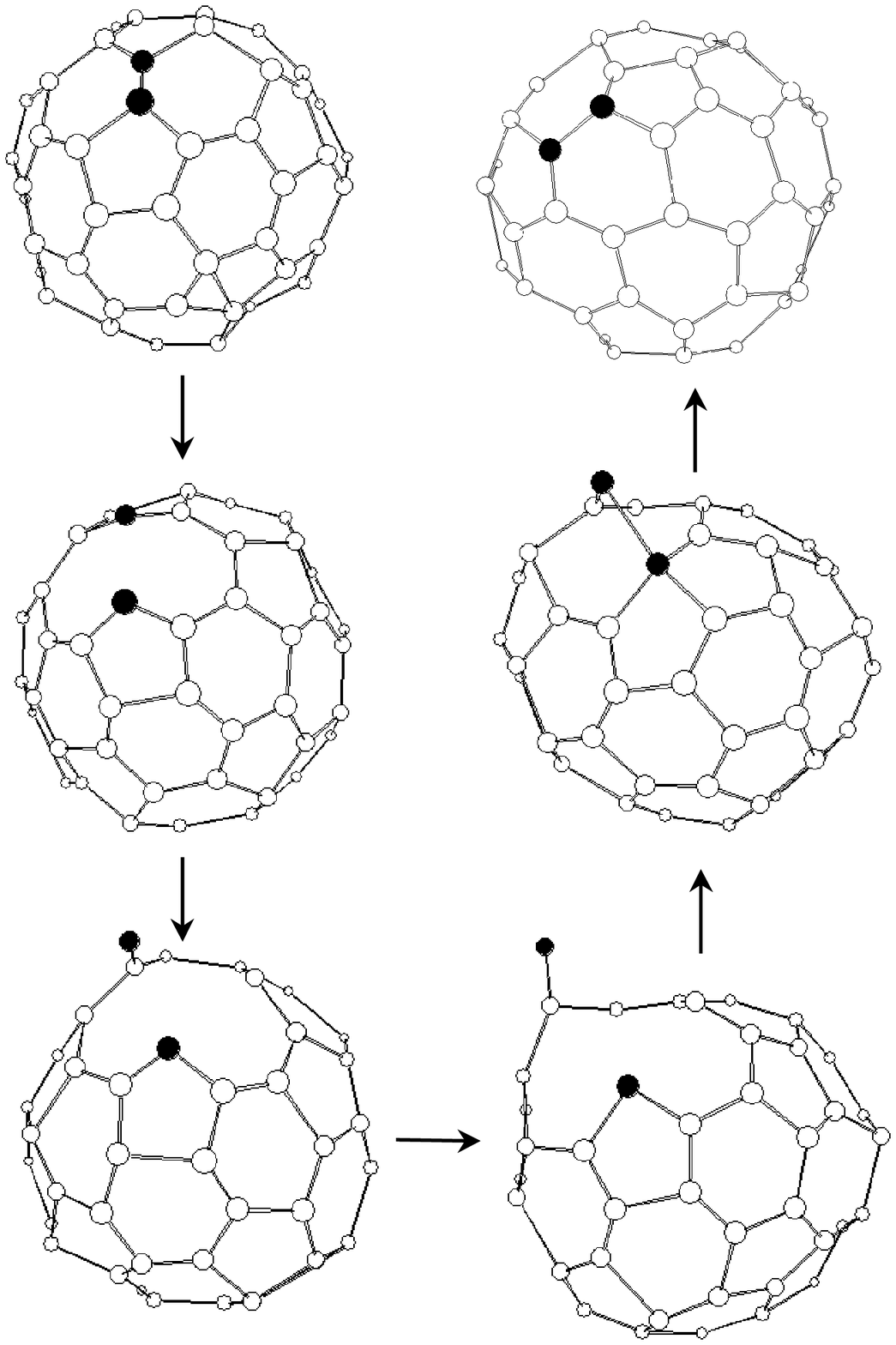}

\vskip 6mm

Fig. 5. Same as in Fig. 4, but for the transformation via path II. $T_{ini}=4680 \pm 5$ K;
$t^{\prime}=0.471$ ns; $\Delta t =$ 27, 54, 136, 299, and 354 fs, respectively.

\newpage

\includegraphics[width=0.6\textwidth]{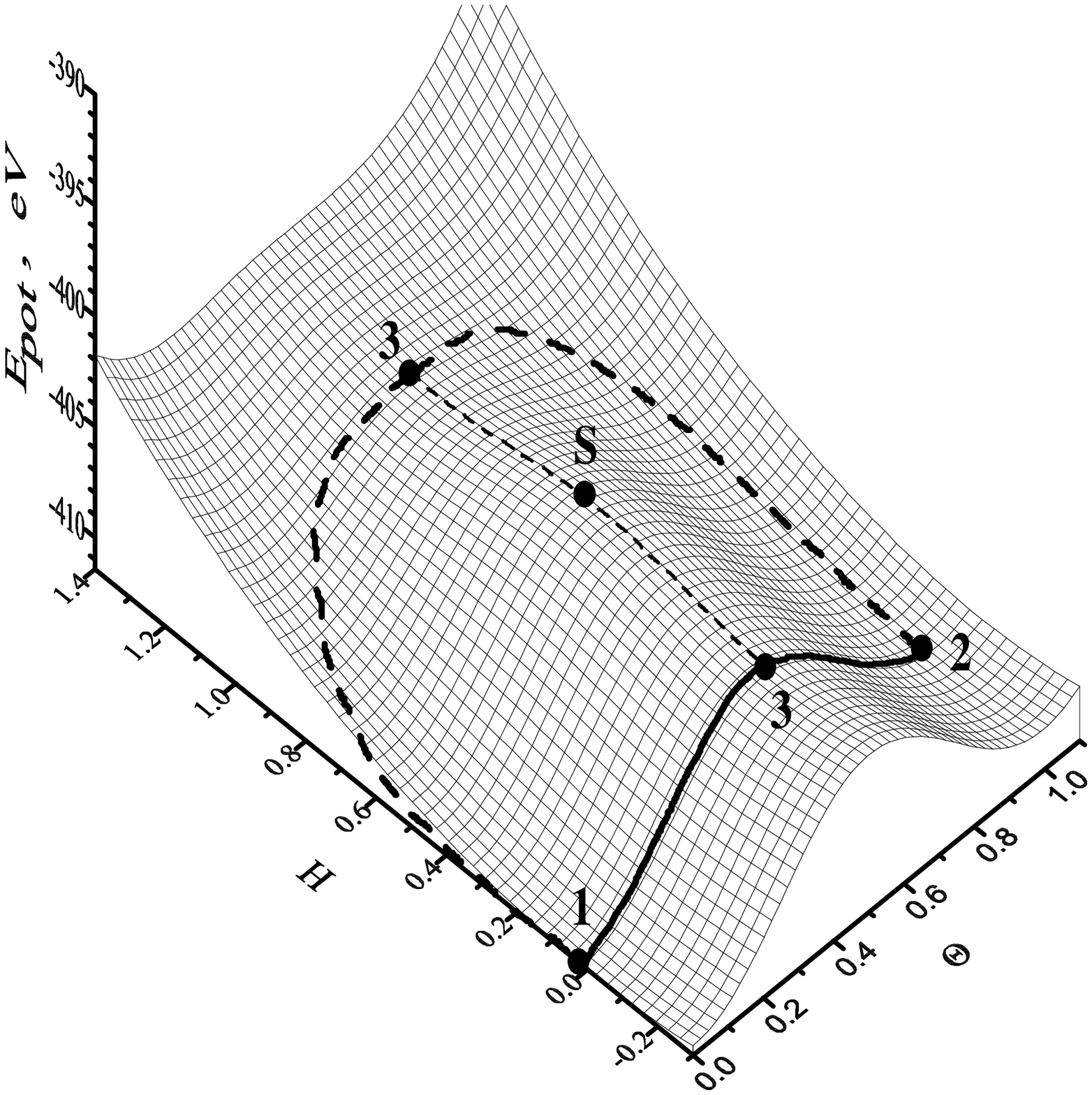}

\vskip 6mm

Fig. 6. Potential energy $E_{pot}$ of the C$_{60}$ cluster vs. two
reaction coordinates $H$ and $\Theta$ in the neighborhoods of (1)
equilibrium and (2) metastable atomic configurations. The energy
of 60 isolated carbon atoms was taken as zero energy. The solid
and dashed lines indicate paths I and II, respectively; {\it 3}
indicates corresponding saddle points; and $S$ is a second-order
stationary point (see the text). The reaction coordinates were
chosen as follows: $H$ is the height through which an atom of the
rotated C-C bond raised over the cluster {}''surface'' normalized
to a maximum height through which this atom rose in its motion
along path II ($H=0$ for path I; a maximum value of $H=1$ was at
the saddle point of path II) and $\Theta$ is the angle of rotation
of the C-C bond normalized to 90$^{0}$ ($\Theta=0$ for
buckminsterfullerene, $\Theta=1$ for the metastable isomer, and
$\Theta \approx 0.5$ for the saddle points and the second-order
stationary point $S$).

\end{document}